# Historical Review of Fluid Antenna and Movable Antenna

Lipeng Zhu[1] and Kai-Kit Wong[2]

[1]Department of Electrical and Computer Engineering, National University of Singapore, Singapore 117583.
[2]Department of Electronic and Electrical Engineering, University College London, London WC1E 7JE, U.K.

Recently, significant attention has been drawn to the development of two emerging antenna technologies known as "Fluid Antenna" and "Movable Antenna" in wireless communication research community, which is greatly motivated by their unprecedented flexibility and reconfigurability for improving system performance in wireless applications. However, some confusions have also ensued on their nomenclature. In fact, both "Fluid Antenna" and "Movable Antenna" are not newly-made terms, and they have a longstanding presence in the field of antenna technology. This article wishes to provide some clarity on this closely related terminology and help dispel any confusion, concern or even dispute on the appropriate use of their names in the literature. Our hope is to unite researchers and encourage more research endeavours to focus on resolving the technical issues on this topic. This article begins by reviewing the historical evolution of these technologies for fostering a clear understanding of their origins and recent development in the realm of wireless communication. We will conclude this article by commenting on other nomenclatures that have emerged in the recent research of Fluid/Movable Antenna.

*A. Historical Review*

In the antenna community, the terms "Liquid Antenna" and "Fluid Antenna" are used interchangeably. While the term "Liquid Antenna" dated back in as early as 1989 [1], [2], the term "Fluid Antenna" appeared rather more recently first in 2008 in a US patent [3] and subsequently a research paper in [4]. Both terms originally referred to antennas utilizing fluid or liquid dielectrics as electromagnetic radiators and are commonly used in the research of antenna design using soft materials. It was 2020 when the term "Fluid Antenna" was first introduced to the field of wireless communication in [5] where the authors extended the concept to represent "*a radical approach that advocates software-controlled position-flexible shape-flexible antenna*". Following this, a more general and detailed definition of the fluid antenna system (FAS) was presented in [6], which "*represents any software-controllable fluidic, conductive, or dielectric structure that can alter its shape and position to reconfigure the gain, radiation pattern, operating frequency, and other characteristics*". In this context, liquid-based and pixel-based fluid antennas were presented as two typical ways of implementation [7], [8]. Several position-flexible FAS prototypes using liquid-based antennas have been reported in [9], [10]. It is noteworthy that the term "Fluid Antenna" used in [5]-[7] has evolved beyond the original definition in [3]. More specifically, the concept of "Fluid Antenna" is general and not limited to any particular way of implementation but encompasses all types of antennas with flexibility and adaptability in their shape and position. Overview articles that cover various aspects of FAS for wireless communications can be found in [7], [11], [12].

On the other hand, the term "Movable Antenna" can be traced back to a book published in 2008 [13] (Section 17.4), with its prototype having surfaced even earlier in 1999 [14]. This term originally referred to antennas endowed with motion/rotation capability through micro-electromechanical systems (MEMS). Later on, a step motor-enabled movable antenna was developed in 2015 [15], allowing for flexible adjustment of the antenna position in radar systems. Following its original definition, the movable antenna concept was introduced and rigorously investigated for wireless communication in 2022 [16], [18], which is applicable to any wireless systems where the "*positions*

*of movable antennas (MAs) can be flexibly adjusted in a spatial region for improving the channel condition and enhancing the communication performance*". Subsequently, an overview of the main advantages of movable antennas for improving the wireless communication performance was given in [19] in terms of signal power improvement, interference mitigation, flexible beamforming, and spatial multiplexing. It was also pointed out in [19] that in addition to the positioning, the rotation/orientation optimization of movable antennas in 3D space can provide additional degrees of freedom for system design.

Readers may have noticed the convergence of "Fluid Antenna" and "Movable Antenna" in recent wireless communication literature, despite their different origins and possibly different ways of realization in practice. Although they both have a longstanding presence in the field of antenna technology, they have been investigated systematically in wireless communication research only in recent years. Conceptually, "Fluid Antenna" and "Movable Antenna" are indeed identical and share the same mathematical model of flexible antenna positioning [5], [16]. Both can thus be considered as interchangeable terms in the literature from the perspective of antenna movement/flexible positioning, when ignoring specific issues in implementation. They have both been adopted by researchers and Table I provides a shortlist of recent works under their respective names.

**Table I: Historical Review of Fluid Antenna and Movable Antenna**

| Nomenclature | First appear | Representative works in wireless communication |
|---|---|---|
| Liquid/Fluid Antenna | 1989 [1]/2008 [3] | Overview [5], [7], [11], [12], [17] |
| | | Channel Modeling [6], [20]-[23] |
| | | Channel Estimation/Recovery [24]-[28] |
| | | Performance Analysis [6], [20], [23], [31]-[52] |
| | | Antenna Position Optimization [55]-[69] |
| Movable Antenna | 2008 [13] | Overview [19] |
| | | Channel Modeling [16], [53] |
| | | Channel Estimation/Recovery [29], [30] |
| | | Performance Analysis [16], [53], [54] |
| | | Antenna Position Optimization [18], [70]-[83] |
| Others | | Flexible-Position MIMO [84] |
| | | Flexible Antenna [85] |

*B. Literature Review and Other Nomenclatures*

In recent years, a significantly growing number of studies have been conducted on fluid antennas and movable antennas for wireless communications, aiming to unveil their advantages over conventional fixed-position antennas, as summarized in Table I. In addition to the overview articles [5], [7], [11], [12], [17], [19], the other representative works have dealt with their channel modeling [6], [16], [20]-[23], [53], channel estimation/recovery [24]-[30] and performance analysis in various wireless systems [6], [16], [20], [23], [31]-[54]. In particular, the channel models mainly include the spatial correlation channel model [6], [20], [21], [23] and the field-response channel model [16], [18], where the former facilitates the performance analysis for fluid/movable antenna systems to characterize their spatial diversity and multiplexing gains while the latter enables the antenna position/orientation optimization in both continuous regions and discrete sets. Under the above channel models, the designs for antenna positioning can be classified into discrete port selection [55]-[68], [70] and continuous position optimization [18], [69], [71]-[78], respectively. Furthermore, increasing research efforts have been devoted to the FAS or movable-antenna array enhanced beamforming design to meet diverse requirements in wireless communication or sensing systems [38], [39], [45], [51], [58], [59], [64], [66], [68], [70], [72]-[74], [76]-[83]. Considering the drastically varying antenna response speed due to different antenna movement/positioning mechanisms in practice, these technologies can be applied to various communication scenarios, such as those exhibiting slowly-varying channels (in majority of the work), fast-varying channels [55], [56], [60], [63] or those with only statistical knowledge of the channels [69], [76].

More recently, some other relevant nomenclatures have also emerged in the literature, such as flexible-position MIMO [84] and flexible antenna [85], among others. These works also focused on the optimization of antenna position to enhance wireless communication performance and have broadened/diversified the research scope in this field.[1] With the emergence of different names after FAS, it is clear that there is growing interest in antenna position flexibility. We are hopeful that collective efforts from the research community will see this field grow even more for years to come and prepare us to tackle the increasing challenges in future-generation wireless networks.

---
[1] Note that in the antenna community, flexible antenna usually refers to bendable antenna for wearable devices [86].